 \documentclass[preprint,review,12pt]{elsarticle}

\usepackage{graphicx}
\usepackage{hyperref}
\usepackage{epstopdf}
\usepackage{epsfig}
\usepackage{url}
\usepackage{amssymb}
\usepackage{amsmath}
\usepackage{helvet}         
\usepackage{courier}        
\usepackage{makeidx}         
\usepackage{multicol}        
\usepackage[bottom]{footmisc}

\let\stdcaption\caption
\let\caption\stdcaption

\journal{Computational Materials Science}

\begin{document}

\begin{frontmatter}

\title{Molecular dynamics simulation of melting of finite and inifinite size graphene}

\author[Label1]{Lian Ming Huei}
\ead{lianminghuei@gmail.com}

\author[Label1]{Tiem Leong Yoon \corref{cor1}}
\ead{tlyoon@usm.my}

\author[Label1]{Yee Yeen Soon}
\ead{soonyeeyeen@gmail.com}

\author[Label2]{Thong Leng Lim}
\ead{tllim@mmu.my}

\cortext[cor1]{Corresponding author}

\address[Label1]{School of Physics, Universiti Sains Malaysia, 11800 USM, Penang, Malaysia}

\address[Label2]{Faculty of Engineering and Technology, Multimedia University, Jalan Ayer Keroh Lama, 75450 Melaka, Malaysia}

\begin{abstract}
We investigate the melting phenomena of pristine, free-standing infinite and finite size graphene sheets via molecular dynamics simulation using AIREBO potential as implemented in the LAMMPS package. In our simulations, the temperature of the systems under investigation are systematically heated up using two independent heating protocols so that the resultant melting temperatures from both schemes can be checked against each other for consistency. The melting temperature of infinite graphene sheet is obtained by following three independent computational experiments. In the first experiment, we simulate the melting of various finite size graphenes, and then determine the melting temperature of infinite graphene sheet as the temperature at which the finite graphenes asymptotically grow in size. In the second experiment, we simulate the melting of infinite single-wall carbon-nanotubes (SWCNTs) with different radius, and then determine the melting temperature of infinite graphene sheet as the temperature at which the radius of SWCNTs asymptotically grows in size. In the third experiment, we heat up an infinite graphene that is formed by constructing a rectangular supercell which is subjected to periodic boundary condition at it sides. Melting temperature for infinite graphene obtained based on the first approach yields $\sim$~5800 K~$\pm$ 22 K. 
The temperatures obtained from the first approach are regarded as the upper limit for melting temperature of finite graphene. The second approach yield $\sim$~5302 K~$\pm$ 36 K, whereas $\sim$~5355~$\pm$ 140 K from the third. 
\end{abstract}

\begin{keyword}

Free-standing graphene, Molecular Dynamics, Melting of, AIREBO.

\end{keyword}
\end{frontmatter}

\section{Introduction}

\noindent 
Thermal stability for pristine, free-standing graphene constitutes a crucial piece of knowledge for them to be used as functional materials in practical applications. Its melting temperature in particular is a  natural benchmark of the thermal stability for carbon-based nanomaterials. However, melting temperature of an infinite, free-standing graphene is hard to measure experimentally. Most measurements of the melting temperature of graphene were carried out on samples sitting on supporting substrate. 
Measured values of melting temperature of infinitely extended free-standing graphene are not directly available. As a reference, the melting temperature of bulk graphite measured experimentally has a large discrepancy and lies between 4000 or 5000 K \cite{Savvatimskiy:Carbon05}.


As a complementary tool to experimental measurement, melting phenomena of nanostructures nowadays can be routinely simulated via atomistic simulation techniques such as molecular dynamics (MD) or Monte Carlo simulations. However, melting temperature of graphene calculated via atomistic simulations as reported in the literature has not converged to a unique value. The criteria for defining ``melting temperature'' of graphene used by researchers in atomistic simulation is not universal, and varies from one paper to another. In addition, in many simulations reported in the literature, such as in \cite{Zakharchenko:JPCM11}, the melting of graphene does not occur abruptly at a single temperature as in bulk materials. There appear to be a pre-melting stage at a lower temperature where carbon atoms begin to shake off from the main body while it still maintain an overall intact form. Only when the temperature hits a higher value does graphene begin to display a ``spontaneous'' melting, after the occurrence of pre-melting at a lower temperature \cite{Los:PRB15}.

While infinitely extended free-standing graphene receives no contribution from finite size effect on its melting temperature, its finite counterparts do. In other words, the melting temperature of a finite graphene is size-dependent. Most measurements of graphene melting phenomena were conducted on supported substrates, and it is experimentally difficult to measure an unambiguous melting temperature for free-standing graphene. Quantifying finite size effect experimentally would be even more so. Specific atomistic simulation approaches such as Monte Carlo and Molecular dynamics, which inherently incorporate temperature effect in their algorithms, are most suitable for the task to gain theoretical insight into the thermal stability and melting behavior of graphene.

Lopez {\it et al.} in an earlier paper (in 2005) \cite{Lopez:Carbon05} performed MD simulation to address the relative thermal stability between finite length carbon nanotube (CNT) and the corresponding nanostrips obtained from opening up these CNT. However, \cite{Lopez:Carbon05} only estimated the bond-breaking temperature for these nanostripts instead of their melting temperature. The bond-breaking temperature, which was abstracted from the caloric curves, represents the lower bound of the graphene melting temperature. They found that the bond-breaking temperature has a non-monotonous size-dependence, ranging from $\sim$2350 K to $\sim$2650 K for nanostrips derived by opening up (2,2) and (3,3) CNT with 3 to 9 times the unit cell length.

In their 2007 MD simulation, in which EDIP forcefield was employed, 
Kowaki {\it et al.} \cite{Kowaki:JPCM07} used a different approach to determine melting temperature of infinite size graphene. They first determined the melting temperature for different radius of infinite single-walled carbon nanotube (SWCNT). As strain energy reduces when radius of infinite SWCNT increases, melting temperature of infinite SWCNT will approach that of infinite size graphene when the radius of infinite SWCNT approaches infinity. The abrupt change in the oscillation pattern of  radial distribution function due to temperature increment, the tale-telling  temperature dependence of mean-square displacement and the dynamic atomic configuration as visually inspected were among the criterion used to determine the melting temperature of infinite SWCNT. They concluded that the melting temperature of an infinitely extended free-standing graphene is 5750 K. 

In their 2011 paper, Zakharchenko~\cite{Zakharchenko:JPCM11} performed Monte Carlo simulation for graphene melting using LCBOPII forcefield. Defining their criteria based on a two-dimensional Lindemann type order parameter, they arrived at a melting temperature of about 4900 K. 

Sandeep K. Singh {\it et al.} in their 2013 paper studied the melting of carbon nano-clusters using both classical MD (with REBO potential) and DFTB-MD methods~\cite{SSingh:PRB13}. The carbon nanostructures studied are minimum energy configurations (in the form of nanoflakes) and were comprised of 98, 142, 194, 322 and 1000 atoms. They used distance-fluctuation Lindemann criteria and caloric curve to determine the melting temperature. The melting temperature of graphene nanoflakes increases with the number of carbon atoms, where the melting temperature increases monotonically from 3800 - 4400 K when the size of nanoflakes increases from 98 to 1000 atoms. The melting temperature of infinite graphene was calculated to be 5500 K. 

The most recent work on the melting of graphene comes from Los {\it et al.} \cite{Los:PRB15}, which is a follow-up of their previous paper \cite{Zakharchenko:JPCM11}, adopted an intrinsic definition of graphene melting based on nucleation theory to arrive at a lower melting temperature of 4510 K. 

The varied values of melting temperature of infinite, free-standing graphene as reported in published simulation works can be traced to a few reasons: a precise definition of melting criteria is not consensually adopted; the use of different forcefield generally results in a different melting behavior (e.g., the sublimation mechanisms as well as the critical temperatures at which they occur, as predicted by the SED-REBO potential was much different than REBO \cite{Steele:MSc13}); difference in the details of the simulation procedure, e.g., the manner by which the temperature of the system is varied and how the melting temperature is measured. 

In this paper we wish to contribute to the elucidation of the melting phenomena of infinite and finite size graphene by conducting systematically designed MD calculations. 
To this end, a robust and reliable choice of forcefield  for performing MD simulation has to be made. The forcefield  of  our choice is the Adaptive Intermolecular Reactive Empirical Bond (AIREBO) forcefield \cite{Stuart:JCP00}, an improvement of the well-known Brenner potential \cite{Brenner:JPCM02}. Both forcefields are well-tested for MD simulation of graphene and CNT. 

\section{Methodology}

\noindent

AIREBO potential, which is implemented in the LAMMPS package \cite{LAMMPS}, is used throughout all of our MD simulations in this study. The timestep used is 0.5~fs. NVT ensemble is used in all simulations. To simulate melting, the systems of interest are heated up until melting phenomena occurs. To this end, Nose-Hover thermostat as implemented in the LAMMPS package is used to control the temperature in all of our simulations. There are non-unique ways to heat up a system in a MD simulation. As a matter of principle, the results obtained (in our case, the melting temperature) from different heating protocols should be consistent with each other. In this paper, we adopt two different heating procedures, dubbed direct- and prolonged-heating protocols, to counter-check for consistency in the melting temperatures obtained from a common simulated system. These heating protocols are described in subsections \ref{sub:DHP} and \ref{sub:PHP} respectively. The melting temperature of infinite graphene sheet is obtained from three independent simulation experiments, which are described in subsections \ref{sub:comp1}, \ref{sub:comp2} and \ref{sub:comp3}.


\subsection{Direct heating protocol}\label{sub:DHP}

\begin{figure*}[htb]%
\begin{center}
\includegraphics*[height=2.40in,keepaspectratio=true]{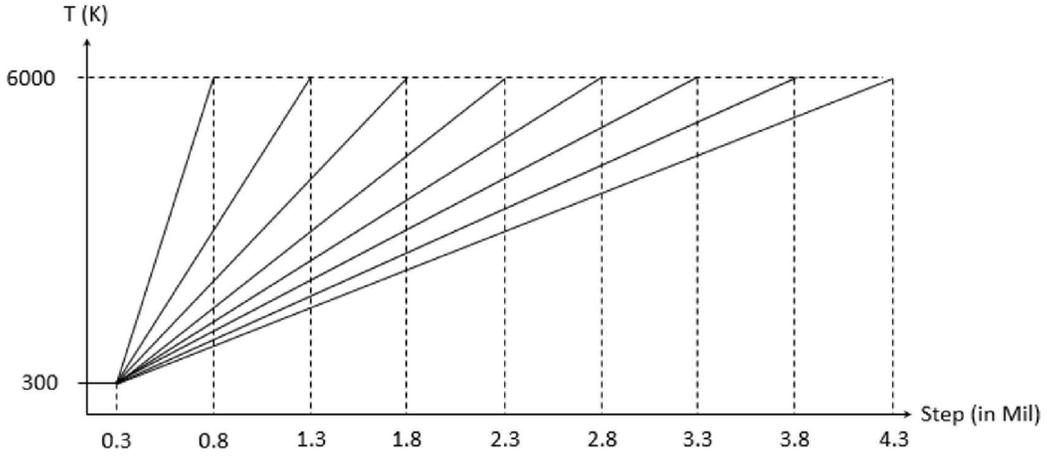}
\caption{Variation in temperature for different steps $N$. Heating rate and $N$ is related via 
$r_N= { (T_{\textrm{target}}-300 {\textrm  K}) \over (N-0.3)}$, $N$ is step in million, 
 $T_{\textrm{target}}$=6000~K in this case.}
\label{Fig3b}
\end{center}
\end{figure*}

The initial configurations of the system to be simulated are first energy-minimized by 
iteratively adjusting the atomic coordinates to achieve local potential energy 
minimum. After that, the systems are equilibrated at 300 K for 300 000 steps (0.15 
ns). Then they are heated to a common temperature $T_{\textrm{target}}=6000$~K but at different rates $r_N$ determined by the number of steps $N$, where $N=$ 0.5 - 4 million steps, with an interval of 0.5 million steps. The heating rate and steps are simply related via 
$r_N= { (T_{\textrm{target}}-300 {\textrm  K}) \over (N-0.3)}$, $N$ is step in million, since there is an initial 300,000 step at $T$=300 K for equilibration. 
The coordinates during the MD are saved at every 2000 steps (1 ps) for post-processing. Figure~\ref{Fig3b} illustrates the 
variation of temperature $T$ of the system up to $T_{\textrm{target}}$ for different $r$. 
The temporal evolution in the MD simulation of a given graphene sheet is 
followed at each heating rate. In particular, the temperature at which melting 
occurs at a fixed heating rate is monitored. For cases where melting did not 
occur up to 6000~K, $T_{\textrm{target}}$ is set to 7000~K. Following 
this procedure, the heating rate dependence of the melting temperature could be determined. In general, the slower the heating rate is the 
longer it takes to achieve the final target temperature, hence more time is 
available for better equilibration. At any finite heating rate, overheating is 
expected. Usually, the larger the heating rate the higher is the melting 
temperature, such as that shown in Fig.~\ref{Fig:pevsstep}(d). Ercolessi 
\cite{Ercoless:ICTP97} discussed about the consequence of estimating the melting 
temperature of a substance by increasing the temperature until the caloric curve 
exhibits a jump. It was commented that the temperature where the jump occurs is 
usually higher than the true melting temperature due to the lacking of liquid 
seed. Hence the melting temperature estimated using this procedure should be 
regarded as an upper bound. 

The ultimate melting temperature in principle should be independent of heating rate. Melting temperatures obtained at a finite heating rate should converge to a unique value in the limit of vanishing heating rate (or equivalently, $N \rightarrow \infty$). This can be achieved by numerical extrapolation. In the present work, the heating-rate-independent melting temperature is derived by
extrapolate 
the data points of the rate-dependent melting temperatures in Fig.~\ref{Fig:pevsstep}(b) to $N \to \infty$, i.e., 
$T_m = \lim_{N \to \infty} T_N$, where $T_N$ refers to the melting temperature of a finite sheet heated with rate $r_N$.
We commented that many graphene melting simulations  reported in the literature either did not report the investigation of the 
convergence of the melting temperature against heating rate (e.g., 
\cite{Lopez:Carbon05,Kowaki:JPCM07}), or just allowed the melting temperature to 
vary with heating rate (e.g., \cite{Zakharchenko:JPCM11}). Such practice could 
potentially render some degree of arbitrariness in their reported melting 
temperatures.

\subsection{Prolonged-heating protocol} \label{sub:PHP}
\noindent

\begin{figure*}[htb]%
\begin{center}
\includegraphics*[height=2.40in, 
keepaspectratio=true]
{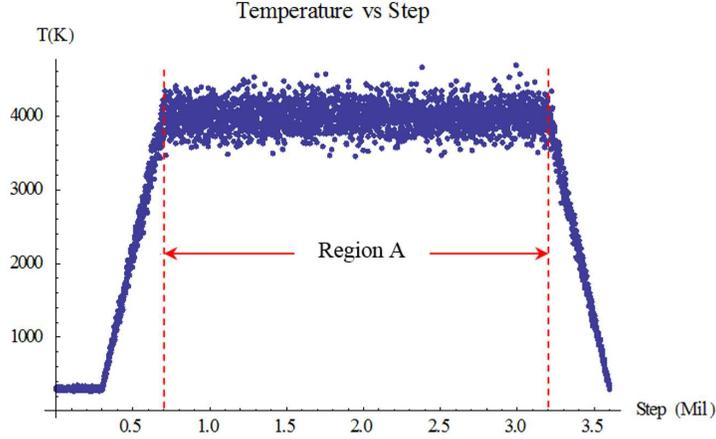}
\caption{A typical profile of the variation of temperature $T$ throughout the MD simulation period for melting of carbon atom configurations with target temperature $T_{\text{target}}$ of 4000~K, using prolonged-heating protocol.}
\label{Fig.3.2}
\end{center}
\end{figure*}

This heating procedure is free from the choices of heating rate. It provides a 
complementary check against the simulations done via the direct-heating protocol. Systems to be investigated are first energy-minimized and then equilibrated at 
300~K for 300~000 steps. The temperature of the systems are then rapidly ramped 
up 
to a predetermined target temperature $T_{\text{target}}$, at which they are equilibrated for a sufficiently lengthy period of time. At the end of the lengthy equilibration, the temperature is rapidly quenched to 300 K at approximately 1 K per 100 steps. The process is repeated for a set of $T_{\text{target}}$ ranging from 4000~K to 7000~K at an interval of 50~K. At each $T_{\text{target}}$ the temporal evolution of the system is monitored for any occurrence of melting phenomena. The length of the equilibration at the $T_{\text{target}}$ plateau, $t_\text{plateau}$, is carefully chosen such that all systems being simulated would either completely melted or remain intact at the end of the equilibration. By this way we can straight forwardly tell whether a system melts, or otherwise, at a fixed $T_{\text{target}}$. The lengthiest equilibration period lasted for 5 ns. Since in this approach $T_{\text{target}}$ is usually set to be very high, the averaged fluctuation in the temperature $T$ during the equilibration plateau is also quite large (approximately in the range of 50~K - 100~K), rendering it not feasible to set a temperature resolution of less than 50~K. 
Fig.~\ref{Fig.3.2} shows an example of temperature profile throughout the simulation for target temperature of 4000~K. 

\subsubsection{Determination of melting temperatures with direct-heating protocol} \label{mtg}

\noindent To determine the melting temperature of a system heated using direct-heating protocol, two quantitative indicators obtainable in the simulations, namely, ({\it i}) the rate of change in the potential energy $PE$, and ({\it ii}) heat capacity $C_V$, both as a function of step, are monitored at a heating rate $r_N$.

\begin{enumerate}[(\it i)]
\item 
To obtain melting temperature by monitoring the potential energy $PE$, its average for 
every three steps is calculated. Melting temperature is taken as the temperature 
at which the maximum rate of change in $PE$ occurs. A typical variation of $PE$ a 
function of step at a fixed heating rate $r_N$ is shown in Fig.~\ref{Fig:pevsstep}(a), 
where \textit{S}${}_{m}$ is the step where the sharpest transition in $PE$ 
occurs. The full width half maximum (FWHM) for the rate of change of $PE$ is 
identified (see Fig.~\ref{Fig:pevsstep}(c)). Fig.~\ref{Fig:pevsstep}(b) shows 
the variation of temperature as a function of step. It is seen that as the 
temperature $T$ grows, so is the fluctuation in $T$. Data points in the $T$ vs. 
step curve that fall in the FWHM is used as samples to determine the average 
and standard deviation of the melting temperature $T_N$ (at the fixed rate $r_N$). If the sample points in the 
FWHM is not symmetric about \textit{S}${}_{m}$, the sample size is enlarged 
until they become so. The sample size must contain a minimum of 31 data to 
justify statistical treatment when calculating variances. The procedure to 
determine $T_N$ is repeated for different heating rate $r_N$, so that a melting 
temperature $T_N$ versus heating rate (in terms of $N$) curve, such as that of 
Fig.~\ref{Fig:pevsstep}(d) is obtained. It is found that, as expected, higher 
heating rate is associated with higher overheating effect and larger uncertainty 
in the value of $T_N$. 
Extrapolation is performed to obtain the value of $T_m$ in the limit of $N \to \infty$ (or equivalently, heating rate $r_N\rightarrow 0$).
This value is taken as the melting temperature for that particular size of graphene. Variance of the melting temperature is calculated by taking into account the contribution of variances of all melting temperatures in Fig.~\ref{Fig:pevsstep}(d). 

\begin{figure}[htb]
\begin{center}
\begin{tabular}{p{3.5in} p{3.5in} } 
\includegraphics*[width=3.02in, height=2.07in, keepaspectratio=false, trim=0.in 0.00in 0.00in 0.00in]{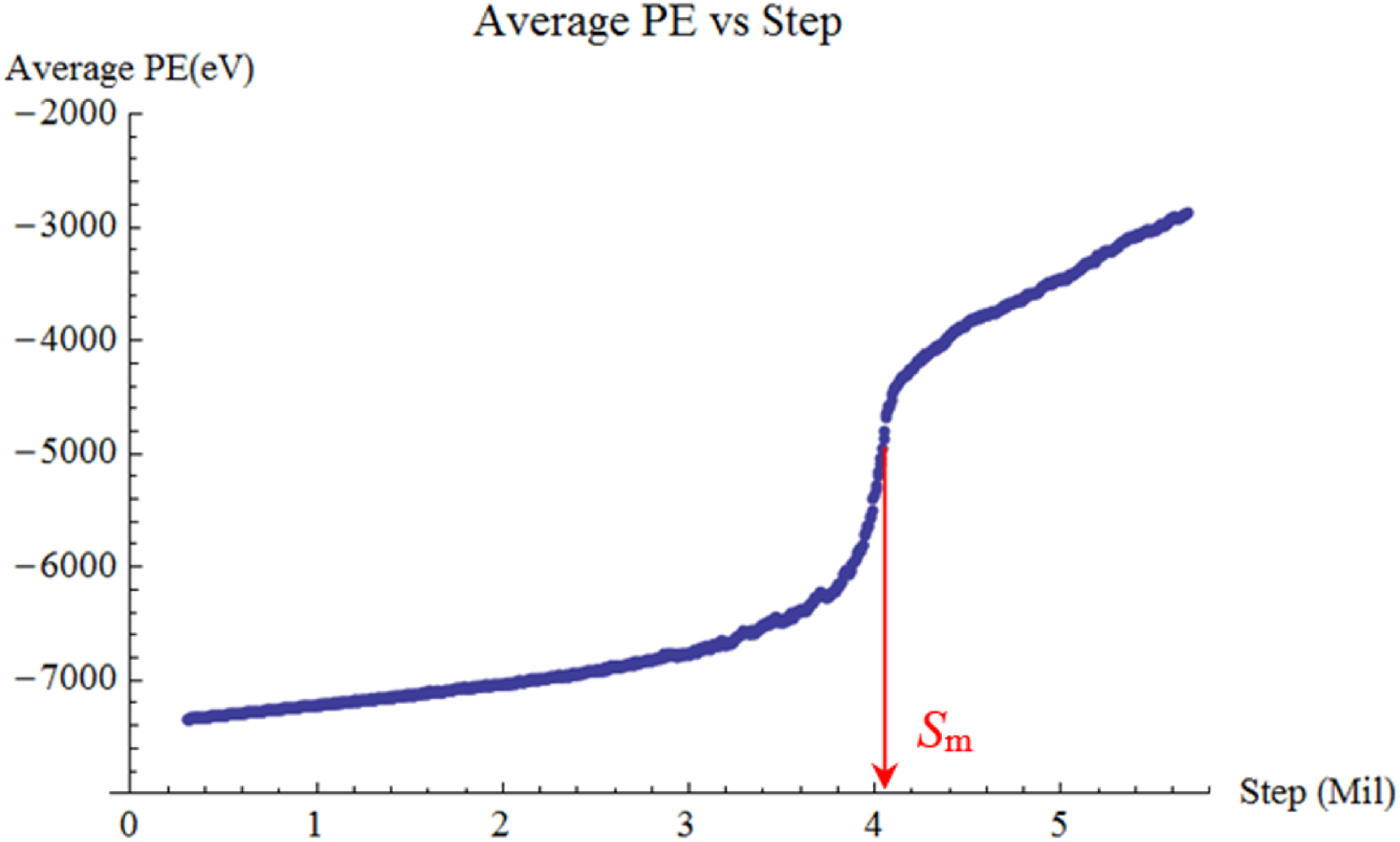} 
& 
\includegraphics*[width=3.02in, height=2.07in, keepaspectratio=false, trim=0.in 0.00in 0.00in 0.00in]{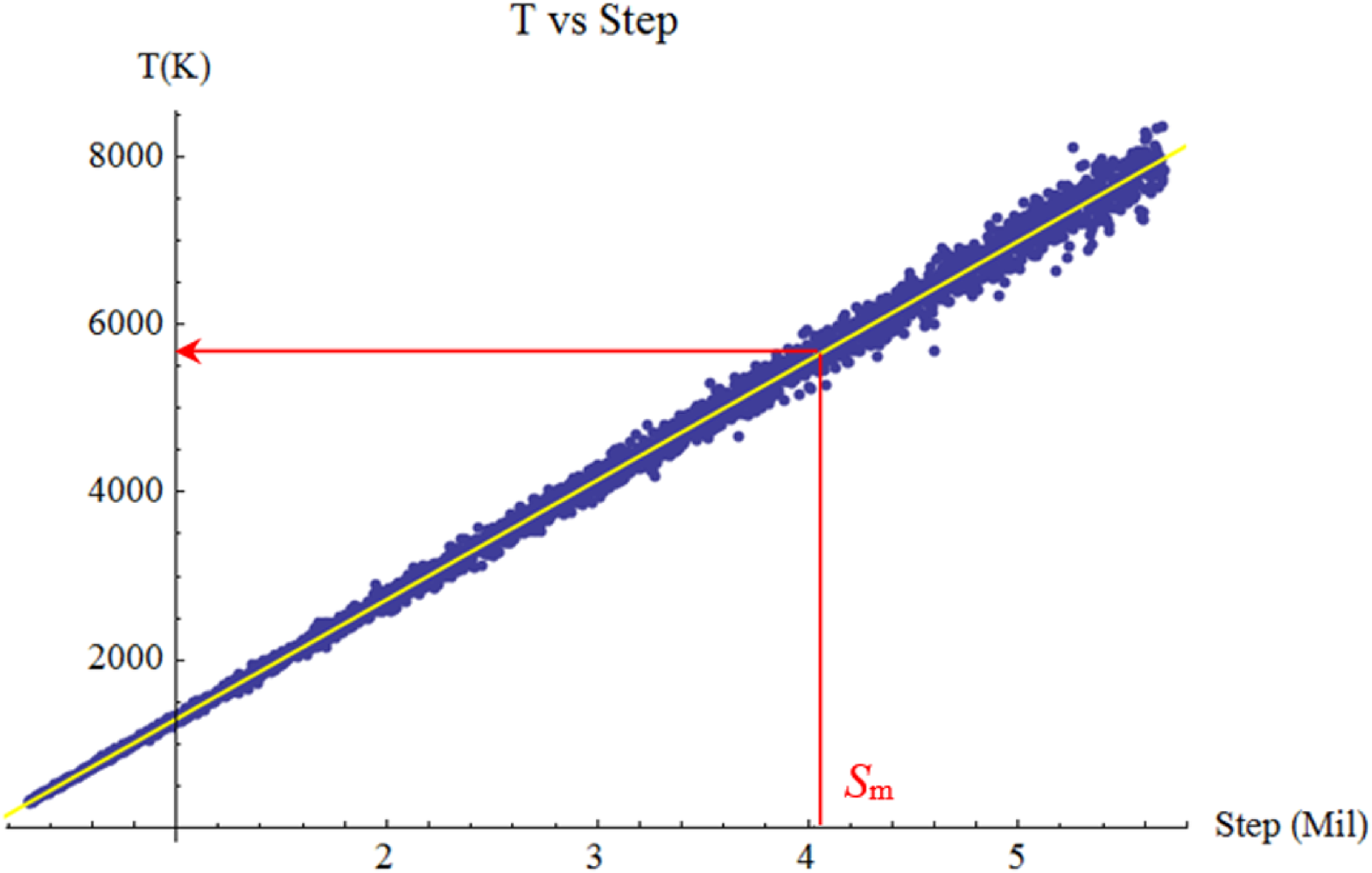} 
\\ (a)\newline  & (b)\newline  
\\
\includegraphics*[width=3.02in, height=2.07in, keepaspectratio=false, trim=0.in 0.00in 0.00in 0.00in]{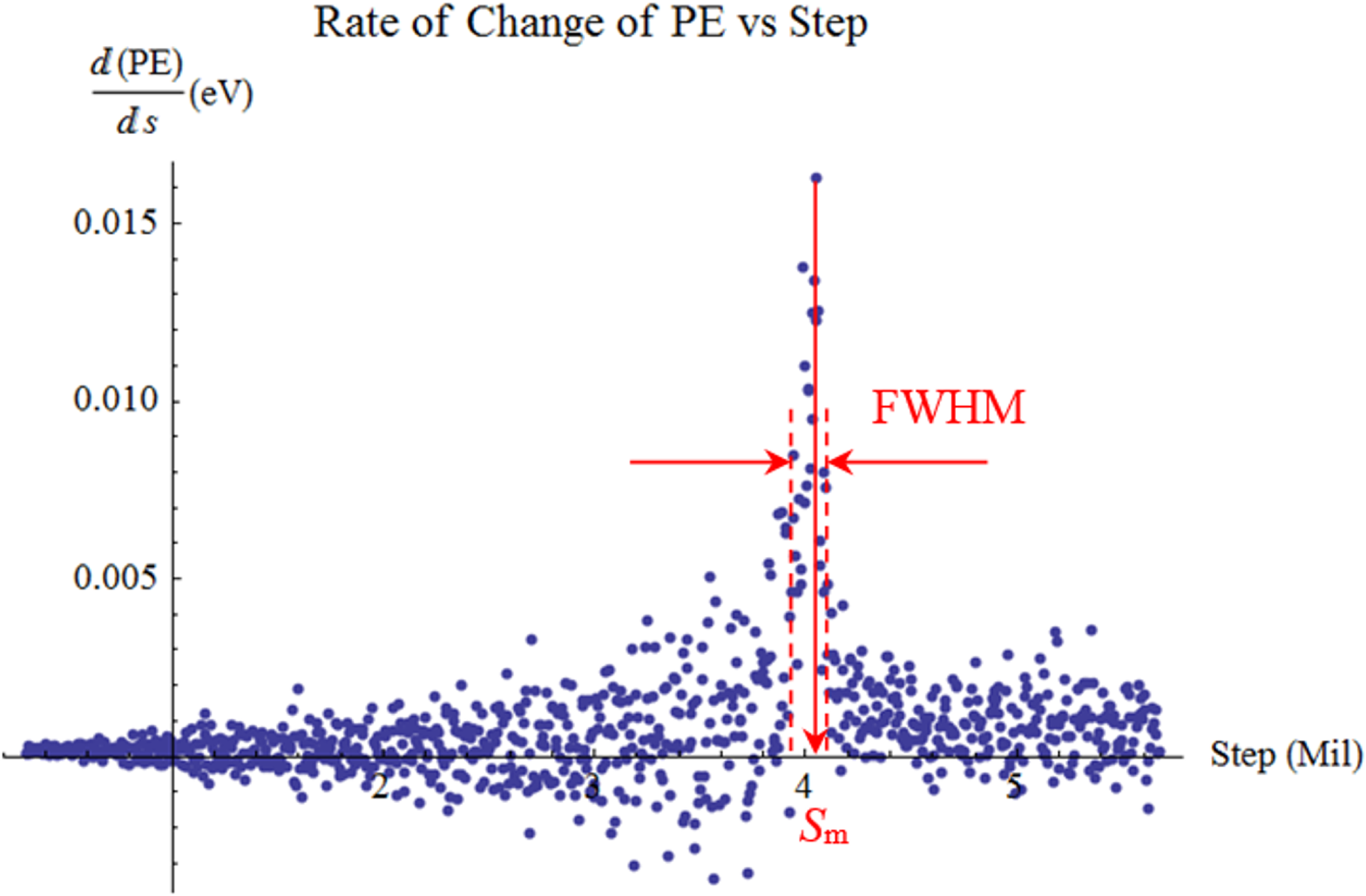} 
& 
\includegraphics*[width=3.02in, height=2.07in, keepaspectratio=false, trim=0.in 0.00in 0.00in 0.00in]{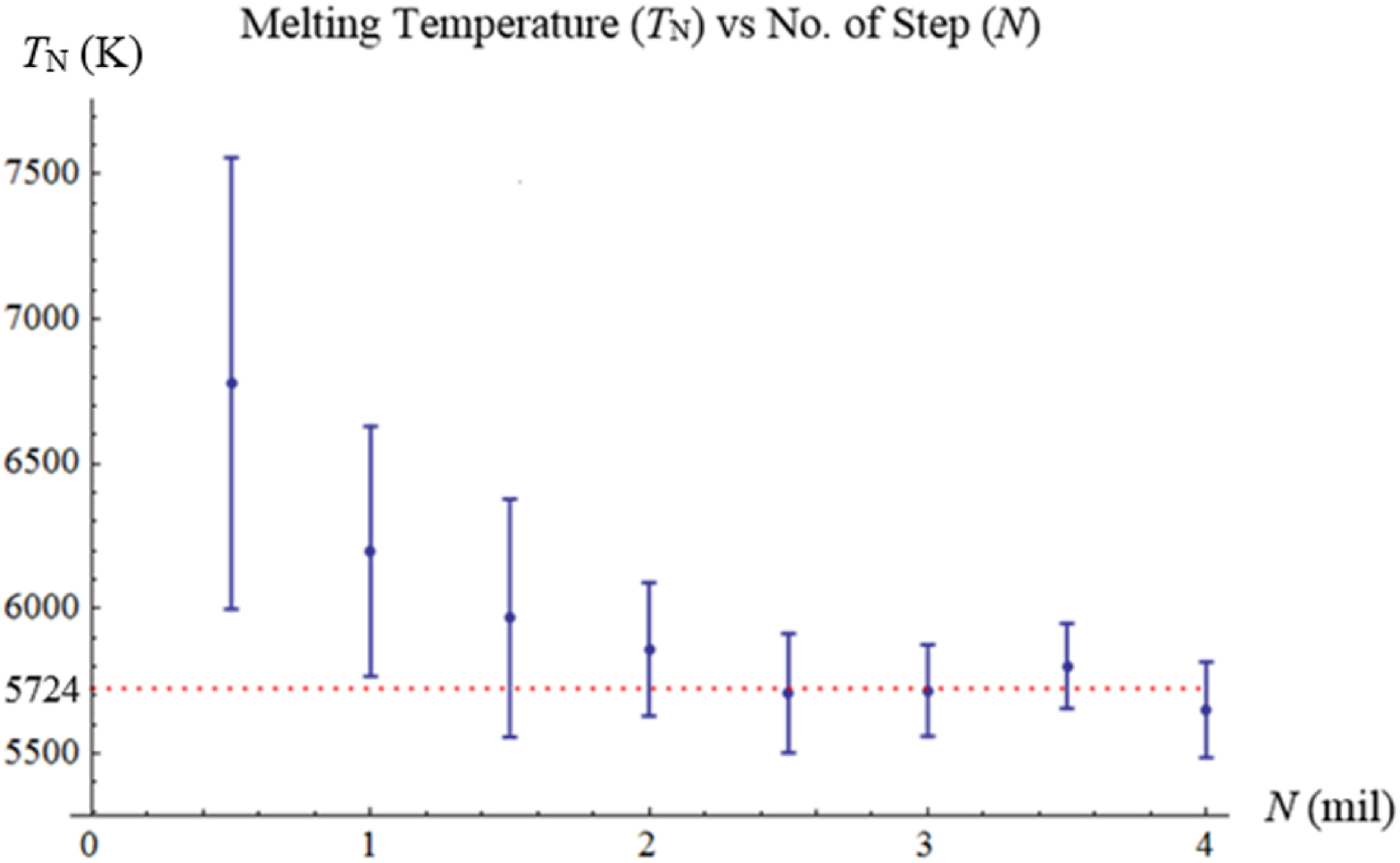} 
\\ (c) & (d) 
\end{tabular}
\caption{Determination of melting temperature of finite graphene sheet comprised of 1024 carbon atom based on the potential energy of the system heated via direct-heating protocol: (a) Average PE versus step at a fixed rate $r_N$. (b) Temperature versus step at a fixed rate $r_N$. (c) Rate of change of potential energy $PE$ versus step at fixed rate $r_N$. (d) Melting temperature $T_N$ versus $N$, where heating rate $r_N$ is inversely proportional to $N$. The red, dotted horizontal line in (d) dictates the value of $T_m$ as obtained from the limit 
$T_m = \lim_{N \to \infty} T_N$.} 
\label{Fig:pevsstep}
\end{center}
\end{figure}

\item $T_m$ can also be obtained independently by monitoring $C_V$. The heat capacity is calculated using the formula 
\begin{equation} \label{GrindEQ__3_3_} 
C_V=\frac{\langle E^{2} \rangle -\langle E\rangle ^{2} }{k_{{\rm B}} T^{2} },  
\end{equation} 
where \textit{E} is the total energy. Note that the numerator measures the variance of energy which is calculated using 25 consecutive data points. The temperature $T$ that appears in denominator in Eq.~\ref{GrindEQ__3_3_} is calculated as the average of $T$ based on the same set of data points. The corresponding temperature at the sharp peak of the $C_V$ versus step curve, such as that shown in Fig.~\ref{Fig:cvvsstep}(a), is taken as the melting temperature.  
\begin{figure}[htb]
\begin{center}
\begin{tabular}{p{3.5in} p{3.5in} } 
\includegraphics*[width=3.02in, height=2.07in, keepaspectratio=false, trim=0.in 0.00in 0.00in 0.00in]{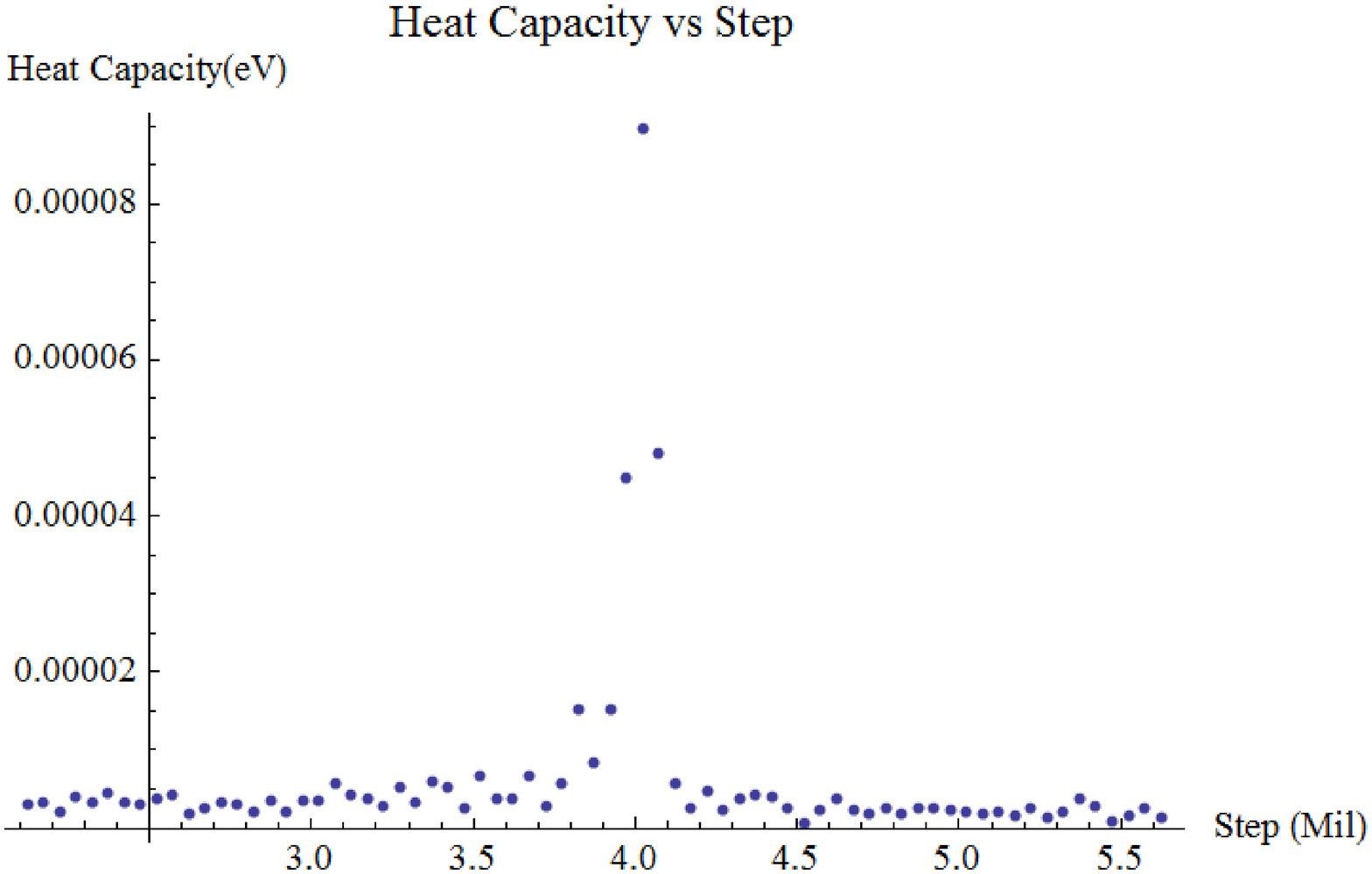} 
& 
\includegraphics*[width=3.02in, height=2.07in, keepaspectratio=false, trim=0.in 0.00in 0.00in 0.00in]{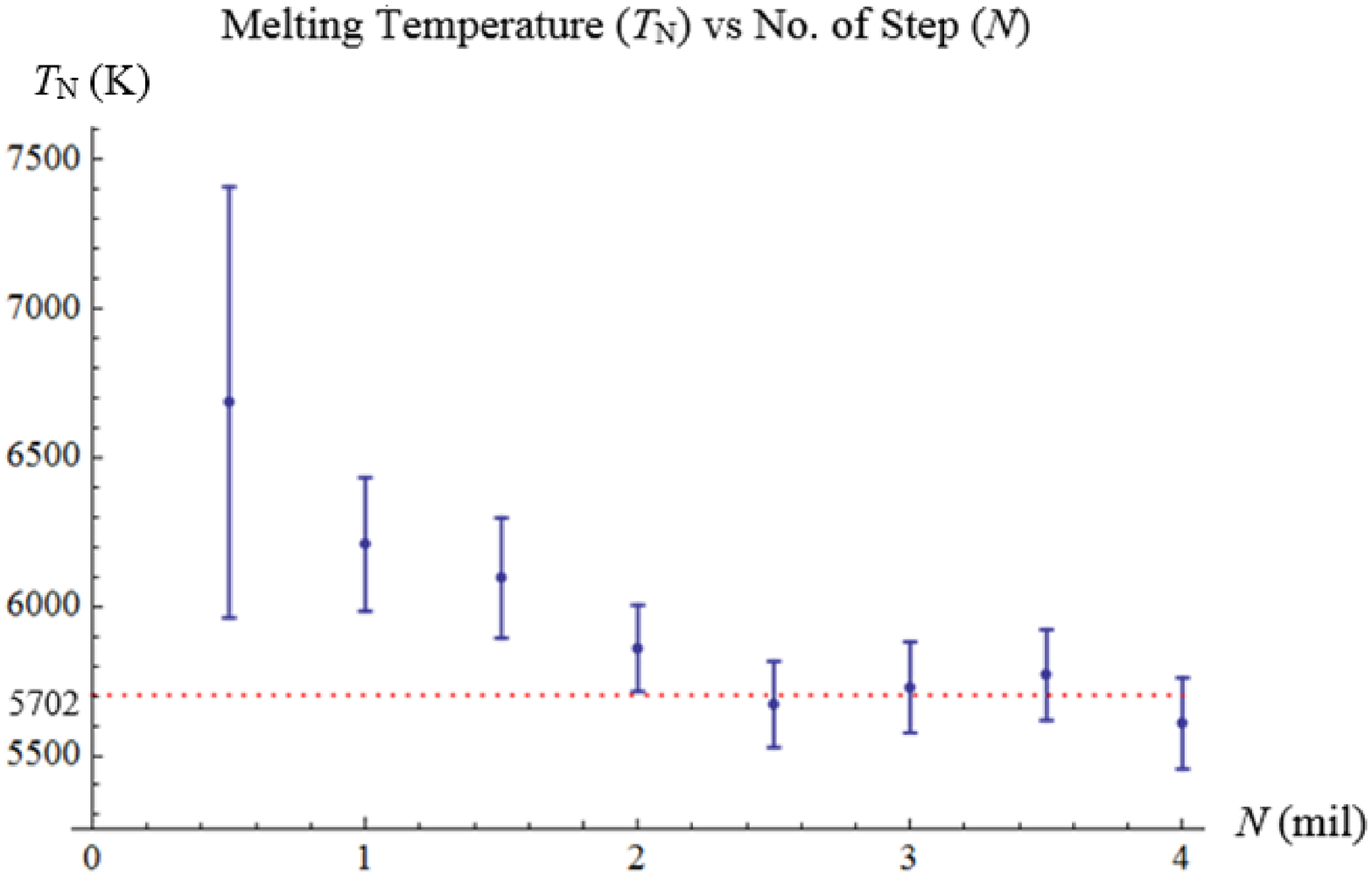} 
\\ (a)\newline  & (b)\newline  
\end{tabular}
\caption{Determination of melting temperature of finite graphene sheet comprised of 1024 carbon atom based on the heat capacity of the system heated via direct-heating protocol: (a) Averaged capacity versus step,  $N$. (b) Melting temperature versus number of step. Heating rate $r_N$ is inversely proportional to $N$. Red dotted line dictates a linear fit through the data points to identify the value of $T_m$ in the limit $r \rightarrow 0$.}
\label{Fig:cvvsstep}
\end{center}
\end{figure}

\end{enumerate}


\subsubsection{Determination of melting temperature with prolonged-heating protocol}
\begin{figure}[htb]
\begin{center}
\includegraphics*[width=6.04in, height=2.07in, keepaspectratio=false, trim=0.in 0.00in 0.00in 0.00in]{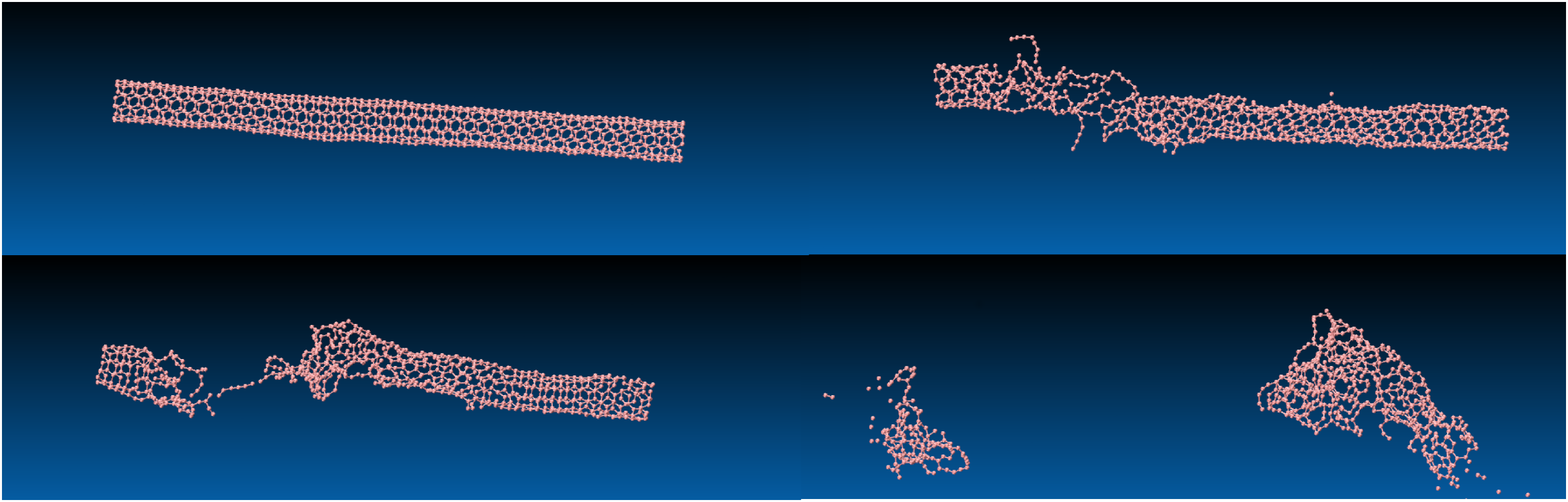} 
\caption{Snapshots of (5, 5) SWCNT with $L{}_{z}$ = 10 nm during a prolonged-heating process at fixed $T_{\text{target}} = 4700$ K.
These snapshots display the SWCNT before, during and after full thermal disintegration as observed in the MD simulation.}
\label{Fig.3.5a}
\end{center}
\end{figure}

\begin{figure}[htb]
\begin{center}
\includegraphics*[width=4.53in, height=3.105in, keepaspectratio=false, trim=0.in 0.00in 0.00in 0.00in]{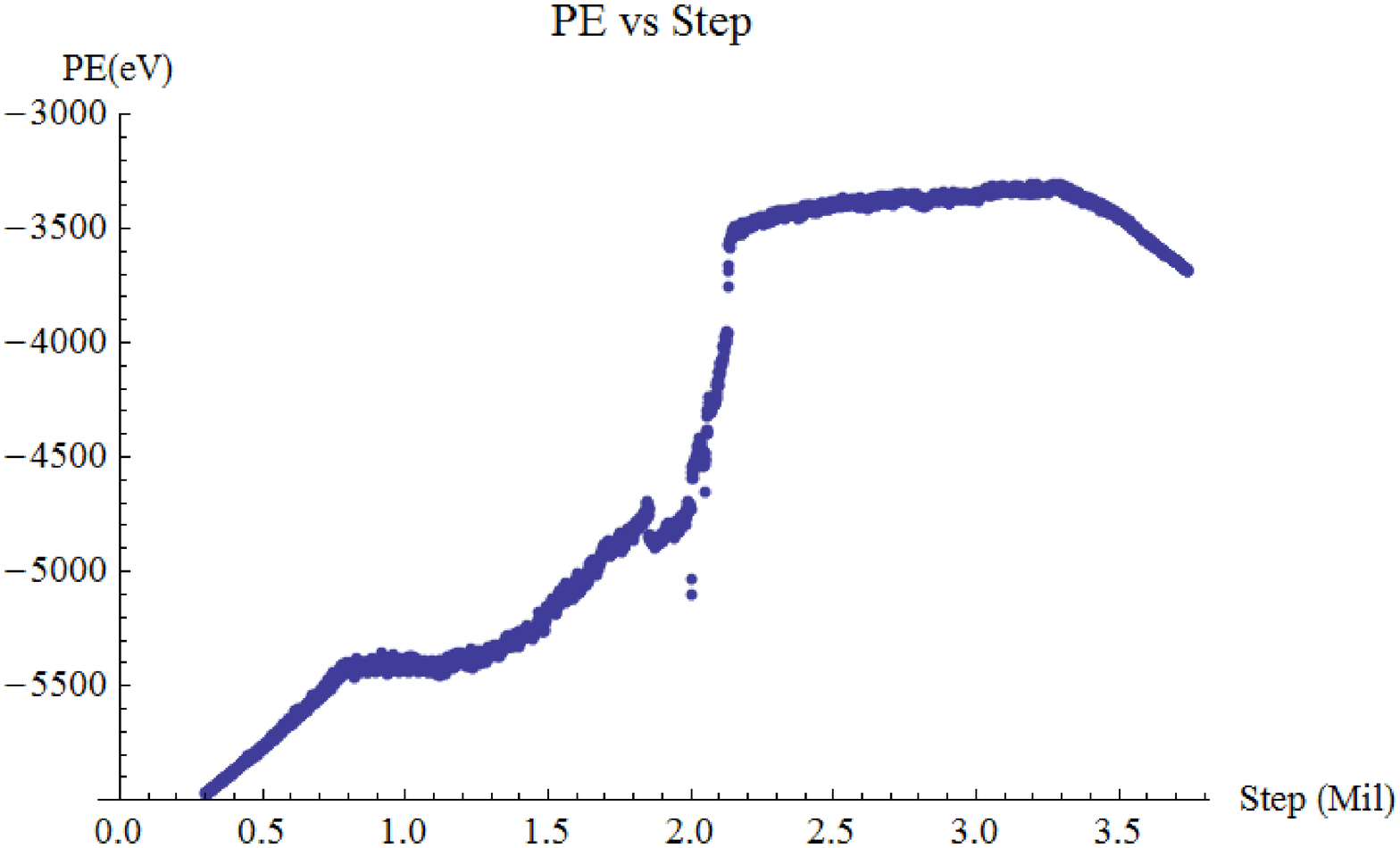} 
\caption{Graph of PE vs step graph in a prolonged-heating process at $T_{\text{target}} = 4700$ K. An abrupt rise of the curve at the step around 2.0 million indicates occurrence 
of structural transition.}
\label{Fig.3.5b}
\end{center}
\end{figure}

\noindent
In the prolonged-heating protocol, it is comparatively straight forward to determine whether at a given target temperature ($T_{\text{target}}$) a system would melt or remain intact. Quantitatively, visual monitoring of the temporal evolution of the simulated system at a fixed $T_{\text{target}}$ provides a convenient way to  confirm the melting scenario. The melting temperature is the lowest $T_{\text{target}}$ at which the originally intact system is visually observed to fully disintegrate while being equilibrated at the $T_{\text{target}}$ plateau. As an illustration, Fig.~\ref{Fig.3.5a} displays a few snapshots taken at a fixed $T_{\text{target}} = 4700$K, in which an infinite CNT disintegrates in consecutive steps. At the same target temperature, an abrupt change in the $PE$ vs. step graph would also show up (illustrated in Fig.~\ref{Fig.3.5b}), indicating transition in the structure is occurring at the corresponding steps. Standard deviation for $T_m$ 
is calculated based on data from the plateau $t_{\text {plateau}}$ (Region A in Fig.~\ref{Fig.3.2}) by excluding outliers. 

\subsection{Melting of finite graphene sheets}\label{sub:comp1}
\begin{figure*}[htb]%
\begin{center}
\includegraphics*[height=7cm,keepaspectratio=true]{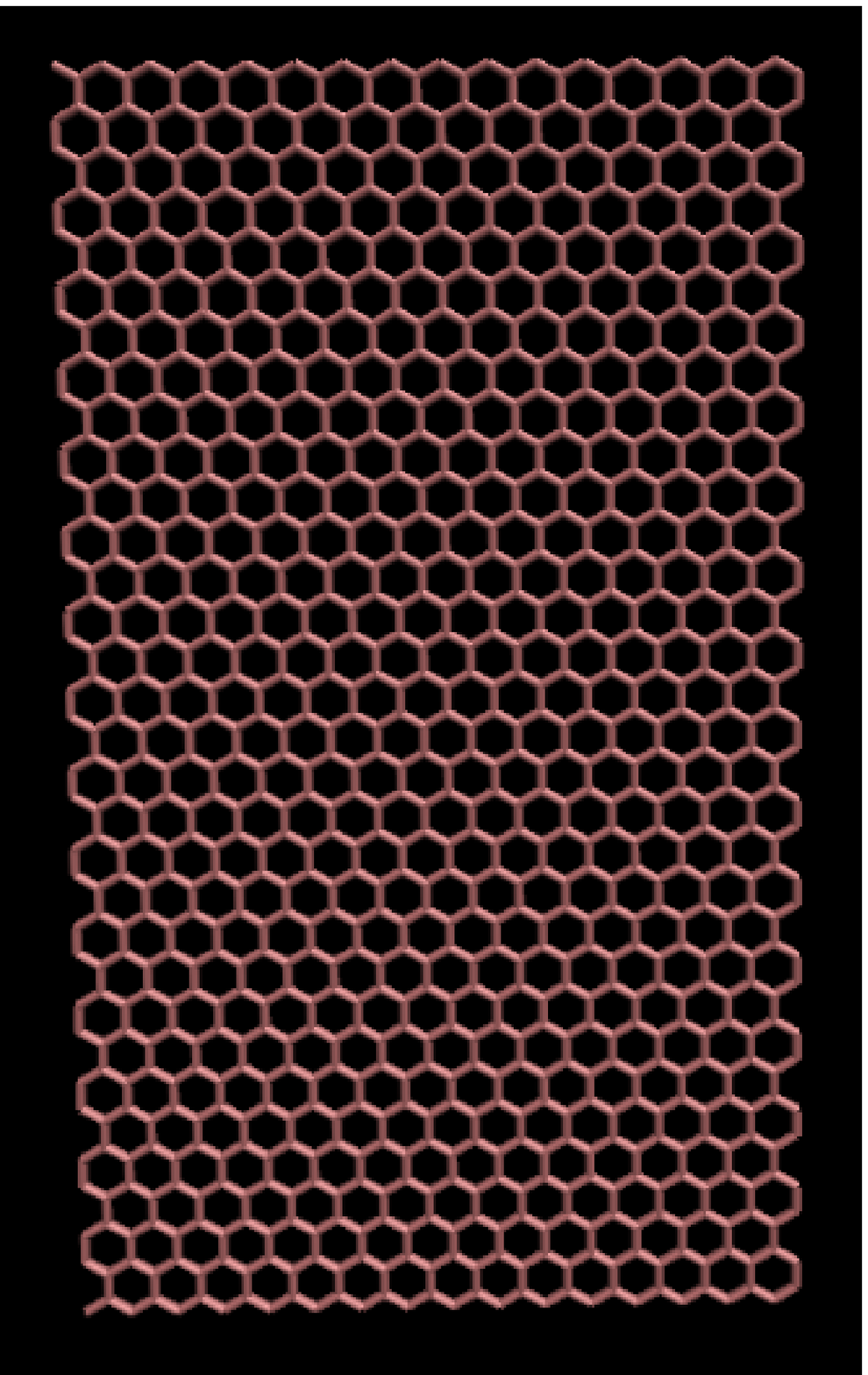}
\caption{Finite graphene sheet of $N$=1024 carbon atoms.}
\label{Fig3a}
\end{center}
\end{figure*}

\noindent
In the first experiment designed to obtain the melting temperature of a free-standing infinite graphene sheet, initial configurations of finite size graphene sheets, such as that shown in Fig.~\ref{Fig3a}, are constructed, with an initial bond length of 0.142 nm which is the bond length of graphite.
The number of carbon atoms $C_N$ in these graphene sheets varies from 256 to 4900. A total of 15 finite sizes have been simulated. The finite graphene sheets are placed in the middle of a simulation box with a size of 50 nm $\times$ 50 nm $\times$ 50 nm, which is assured to be large enough to contain the largest graphene sheet. Terminated boundary condition at the boundaries of the box is imposed. These graphene sheets are first energy-minimized and then heated via direct-heating protocol to obtain their melting temperatures. This process is repeated for graphene sheets of various sizes $C_N$ so that the melting temperature as a function of graphene size is obtained.

\subsection{Melting temperature of infinite graphene sheet via the Kowaki approach}\label{sub:comp2}
\noindent
In the second experiment, the melting temperature is obtained by using an approach used by Kowaki in Ref.~\cite{Kowaki:JPCM07}. In this approach, an infinitely long SWCNT with radius $R$ (which has an initial bond length of 0.142 nm) was heated up via prolonged-heating protocol to locate its melting temperature, $T_{\text {SWCNT}}$. The infinite length SWCNT is realized by constructing a rectangular supercell of sizes $L_x \times L_y \times L_z$, which is subjected to periodic boundary condition along the axial direction (chosen to be along the $z$-axis) and terminated boundary condition at the box surfaces $L_x \times L_z$ and $L_y \times L_z$. The sides $L_x$ and $L_y$ are set to be large enough (around 20 times the diameter of the SWCNT it contains) to avoid significant boundary effects on the MD final output. Each of these initial CNT configurations is energy-minimized as in the case of the finite graphene sheets. We noted that this bond length may change after energy minimization. To rule out numerical arbitrariness, the simulation results (e.g., the melting temperature) should be independent of the length $L_z$, since periodic boundary condition always generates an infinite CNT along the $z$-direction irrespective of the choice of $L_z$ as long as it exceeds certain minimal length. To ensure this is the case, we perform a convergence test on results of the MD simulation against $L_z$. We simulate the melting of an armchair (5,~5) infinite SWCNT, via prolonged-heating protocol, using different choices of $L_z$. The dependence of melting temperature on $L_z$ is investigated for $L_z$ =\{4 nm, 5 nm, 10 nm, 40 nm\}, so that a suitable choice for $L_z$ could be made for the subsequent MD simulation.
The melting temperatures of infinite SWCNT for different values of radius, $T_{\text {SWCNT}}(R)$, are then obtained by heating these SWCNTs via prolonged-heating protocol. 

Infinite SWCNT is often thought of as folding of infinite graphene. This introduces strain energy in infinite SWCNT which is defined as \cite{Kowaki:JPCM07}:

\begin{equation}\label{PE}
E_{\text{strain}} = P_{\text{SWCNT}} - PE_{\text{graphene}},     
\end{equation}
where $PE$ is potential energy. Melting temperature strongly depends on the strength of the bonds. $PE$ can be considered as negative of binding energy: 
\begin{equation} \label{ES}
E_{\text{strain}} \sim k_{B} T_{\text{SWCNT}} - (-k_B {T}_{\text{graphene}})
\end{equation}
where \textit{k}${}_{B}$ is Boltzmann constant, 
\textit{T}${}_{\text{graphene}}$ is the melting temperature of graphene. Strain energy per atom depends on the radius of infinite SWCNT, \textit{R}:

\begin{equation}
{E}_{\text{strain}}=\frac{C}{R^{2} },
\end{equation}
where \textit{C} is a constant. Combining equation \eqref{PE} and equation \eqref{ES}:
\begin{equation} \label{GrindEQ__2_4_} 
T_{{\text{graphene}}} -T_{{\text{SWCNT}}} \sim \frac{C}{k_{{\rm B}} R^{2} }.  
\end{equation} 
The larger the radius, the lower the strain energy, the closer the melting temperature of infinite SWCNT to that of infinite size graphene. 

The melting temperature of infinite graphene sheet $T_\text{graphene}$ is obtained by 
fitting $T_\text{SWCNT}(R)$ against radius $R$ according to Eq.~\eqref{GrindEQ__2_4_}.
The error associated in estimating this parameter is taken as uncertainty in estimating $T_\text{graphene}$.
The melting temperature of infinite graphene can hence be determined as the limiting case 
\begin{equation}
T_{{\text{graphene}}} = \lim_{R \to \infty} T_{{\text{SWCNT}}}(R).
\end{equation}


\subsection{Melting of infinite graphene sheet formed by periodic supercells}\label{sub:comp3}
\noindent
The third independent experiment to measure the melting temperature of an infinite graphene sheet is done as followed:  
An infinite graphene is formed by constructing a rectangular supercell consists of $N_a$ atoms and of dimensions $L_x \times L_y \times L_z$. The supercell is subjected to periodic boundary condition at the sides $L_x$ and $L_y$, and terminated boundary condition at the box surfaces $L_x \times L_y$. The height of the supercell box along the $z$-axis (i.e, $L_z$) is set to a large value (20 nm) to minimize any possible interaction from neighboring cell in the $z$-direction. Each initial configurations is energy-minimized before subjecting them to a prolonged-heating protocol. The melting temperature $T_m$ for each infinite graphene sheet formed by different choice  of $N_a$ ($N_a$ = \{1024, 1156, 1296, 1444, 1644\}) is obtained. Each periodic structure formed with these supercells in principle effectively mimics an infinite graphene sheet. Since the number of atom, $N_a$, forming these supercells is relatively large, it is expected that the melting point of the resultant infinite graphene sheet to be weakly dependent of $N_a$. 


\section{Result and discussion}

\subsection{Melting Temperature of finite graphene sheets}
\begin{figure}[htb]
\begin{center}
\includegraphics*[width=4.53in, height=3.105in, keepaspectratio=false, trim=0.in 0.00in 0.00in 0.00in]{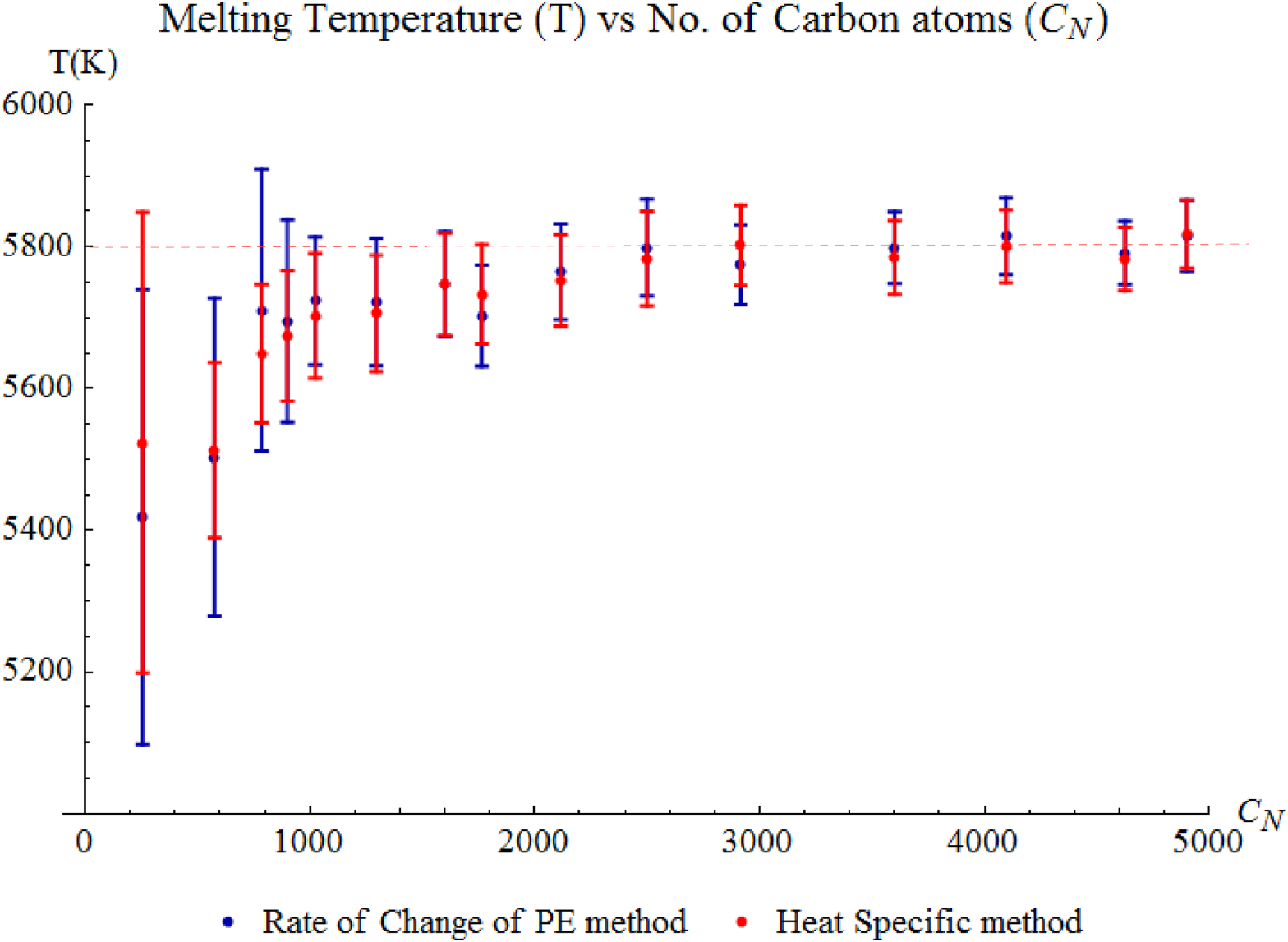}
\caption{Melting temperature for different sizes of finite graphene as determined using maximum rate of change of potential energy (in blue) and heat capacity (in red). These results are obtained based on the procedures as described in \ref{sub:comp1}. The error bars show the standard deviation of melting temperatures at a specific size. Red dotted asymtote indicates the value of $T_m$ as $C_N \rightarrow \infty$.}
\label{Fig.4.1}
\end{center}
\end{figure}

Fig.~\ref{Fig.4.1} shows the melting temperatures determined using potential energy changing rate 
${d\textit{(PE)} \over ds}$ and heat capacity. Melting temperatures determined using both indicators are  consistent to each other. The overall trend shows that the melting temperature increases for $C_N$ starting from 256 until $C_N = 2116$, after which $T_m$ appears flatten asymtotically.
Error bar is especially large at small $C_N$ as it is generally more difficult to peg the temperature at large values for small size system in a MD experiment. The melting temperature as $C_N \rightarrow \infty$ as determined by both indicators estimated as $\sim$~5800~K $\pm$ 22 K (indicated as the asymtote (red dotted line) drawn in Fig.~\ref{Fig.4.1}). 
Fig.~\ref{Fig.4.1} also provides the evidence that melting temperature for graphene sheet smaller than $C_N \sim 2000$ is less than that of an infinite sheet. Within the limits of the error bars in Fig.~\ref{Fig.4.1}, it is evident that melting temperature of finite graphene sheets is size-dependent, especially in the small size region, $C_N \stackrel{<}{\sim} 2000 $. However a more detailed melting temperature dependence on $C_N$ for small graphene sheets cannot be determined to any better precision due to inherent large temperature fluctuation (hence uncertainty) associated with the simulation. 


\subsection{Melting Temperature of SWCNT}
\noindent The melting temperatures of infinite (5,5) SWCNT constructed by different $L_{z}$ are summarized in Table~\ref{Table.4.1}. 
The average melting temperature obtained via prolonged-melting protocol  
is 4625 $\pm$ 103~K. The choice of $L{}_{z}$ does not significantly 
influence the melting temperature obtained.  Hence $L{}_{z}$ = 4 nm is chosen 
for subsequent simulations of infinite SWCNT (of different radius) melting. 
The value 4625 $\pm$ 103~K does not differ much from 4800~K as 
determined by Zhang {\it et al.} using Tersoff potential 
~\cite{Zhang:Nanotechnology07}, but deviates quite significantly from that 
obtained by Kowaki who used EDIP \cite{Kowaki:JPCM07}, which is $\sim$3500~K. 
The difference could be due to the use of different potential in these 
simulations. 

\begin{figure}[htb]
\begin{center}

\includegraphics*[width=4.53in, height=3.105in, keepaspectratio=false]{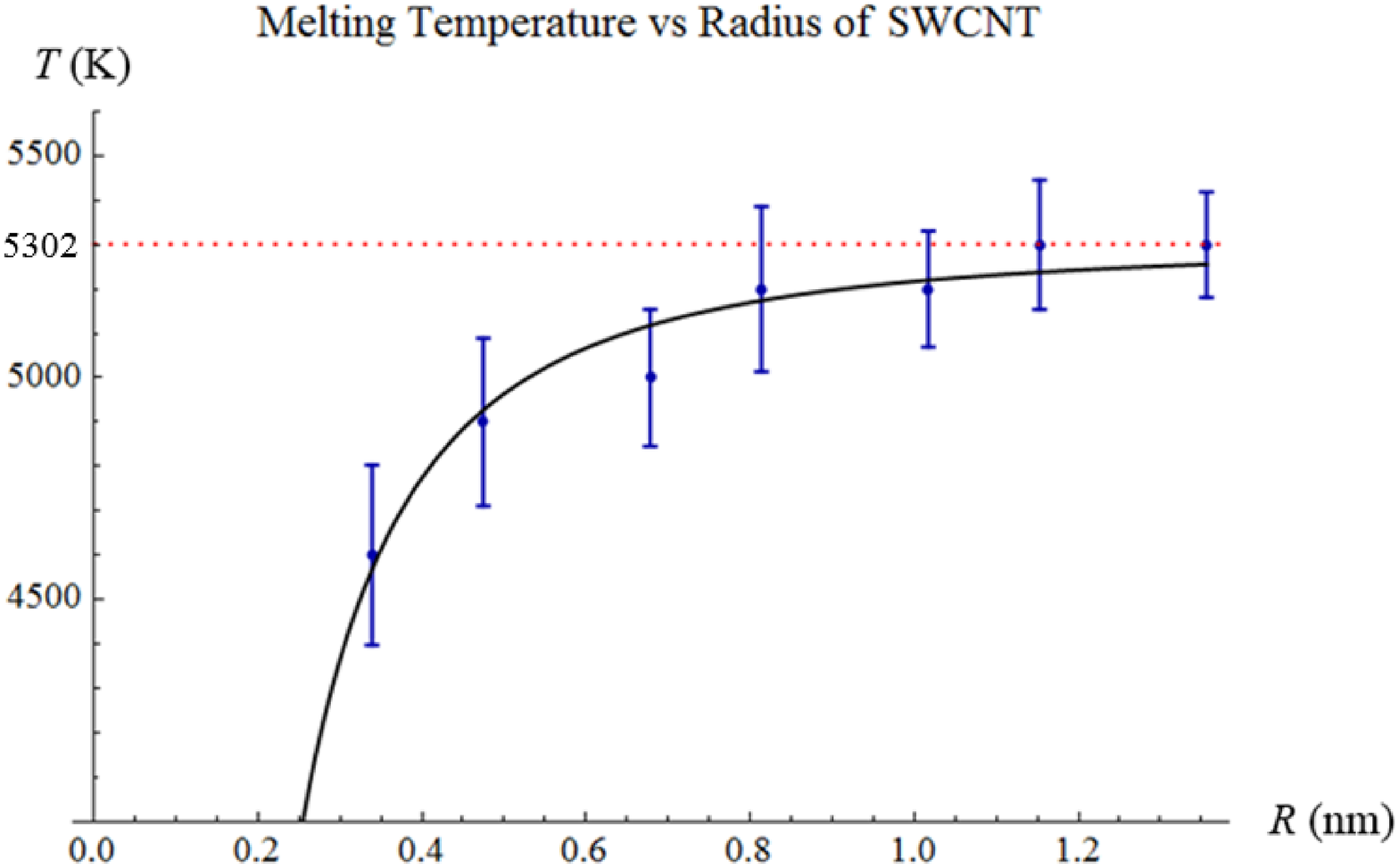}
\caption{Melting temperature of infinite SWCNT as a function of radius.}
\label{Fig.4.2}

\end{center}
\end{figure}

\begin{table}
\caption{Melting temperature and corresponding standard deviation for infinite SWCNT with different $L{}_{z}$.}
\begin{center}
\begin{tabular}{|c|c|c|} \hline 
\eject $L{}_{z}$ (nm) & Melting temperature (K) & Standard deviation (K) \\ \hline 
4 & 4600 & 203 \\ \hline 
5 & 4700 & 215 \\ \hline 
10 & 4700 & 212 \\ \hline 
40 & 4500 & 195 \\ \hline 
\end{tabular}
\end{center}
\label{Table.4.1}
\end{table}
Melting temperature of infinite SWCNT with its radius $R$ is shown in 
Fig.~\ref{Fig.4.2}. After the temperatures are fitted using 
Eq.~\eqref{GrindEQ__2_4_}, the melting temperature of infinite size graphene is 
estimated as 5302 K $\pm$ 36 K (red line in Fig.~\ref{Fig.4.2}). 
Note that the melting temperature of infinite SWCNT converges to 
that of infinite size graphene when the radius $R > 1.15$ nm. The constant $C$ 
in equation \eqref{GrindEQ__2_4_} is estimated as 7.3 meV nm${}^{2}$. The $C$ 
constant for AIREBO potential is at lower end among the $C$ value of all other 
potential as shown in Table~\ref{tab:potential}.

\begin{table}
\caption{$C$ values for different potential.}
\begin{center}
\begin{tabular}{|c|c|} \hline 
Potential & $C$ (meV nm$^{2}$) \\ \hline 
Present Work (AIREBO) & 7.3 \\ \hline 
Tersoff & 15 \cite{Robertson:PRB92} \\ \hline 
Tersoff -- Brenner & 12 \\ \hline 
Density Functional Theory & 19.6 \cite{Hasegawa:PRB06} \\ \hline 
EDIP & 24.7 \cite{Kowaki:JPCM07} \\ \hline 
\end{tabular}
\end{center}
\label{tab:potential}
\end{table}


\subsection{Melting Temperature of infinite graphene sheet via periodic 
supercells}

\begin{figure}[htb]
\begin{center}
\includegraphics*[width=4.53in, height=2.7in, keepaspectratio=false]{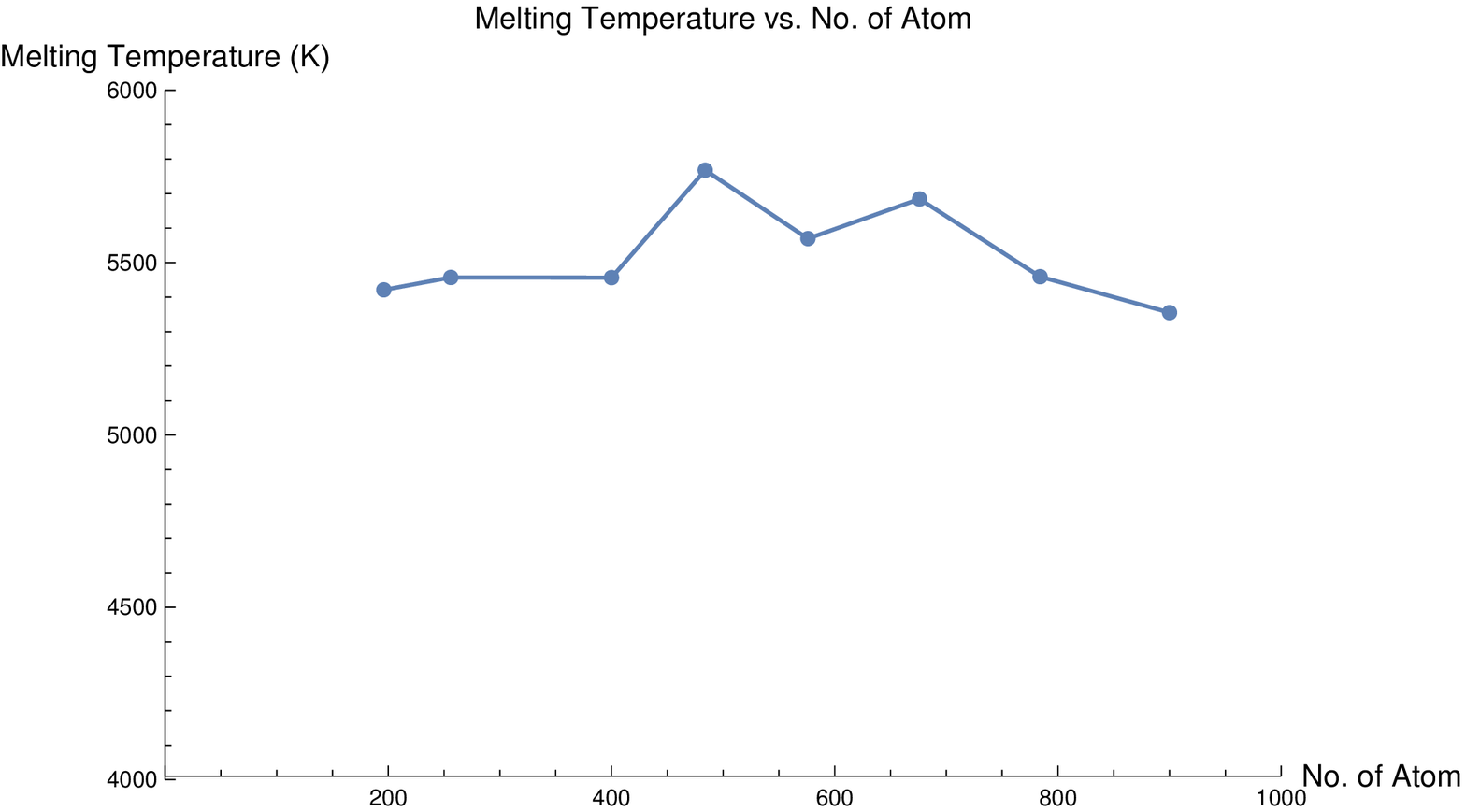}
\end{center}
\caption{The melting temperatures of infinite graphene sheet constructed by 
different supercell size $N_a$.}
\label{Fig.TmvsNa}
\end{figure}

\noindent The melting temperature of infinite graphene sheet constructed by 
different supercell size $N_a$ are summarized in Table~\ref{Table.TmvsNa}, with the 
corresponding graph shown in Fig.~\ref{Fig.TmvsNa}. It is 
found that $T_m$ obtained fluctuates in the range of $\sim 5350$~K - 
$\sim 5770$~K. From Fig.~\ref{Fig.TmvsNa} we estimate the averaged $T_m$ is 5355~K, with estimated standard deviation of 140~K.  

\begin{table}
\caption{Melting temperatures of infinite graphene sheet constructed by 
different supercell size $N_a$.}
\begin{center}
\begin{tabular}{|c|c|c|} \hline 
$N_a$ & Melting temperature (K) \\ \hline 
900	& 5354.975 \\ \hline 
784	& 5459.15 \\ \hline 
676	& 5684.5 \\ \hline 
576	& 5569.5 \\ \hline 
484	& 5768.0 \\ \hline 
400	& 5456.5 \\ \hline 
256	& 5457.0\\ \hline 
196	& 5421.0\\ \hline 
\end{tabular}
\end{center}
\label{Table.TmvsNa}
\end{table}

\section{Conclusion}

\noindent Estimation of melting temperature of free-standing infinite graphene 
 sheet has been attempted by using three independent set of systematically designed  
MD experiments. The commonality of all these experiments is that they all are carried out using the same forcefield, namely, AIREBO. Statistically error bars associated with the obtained melting temperatures are also quantified. 

In the first experiment, in which direct-heating protocol is used, the melting temperature of finite size graphene sheets displays evident size-dependence for size $C_N\stackrel{<}{\sim} 2000$. The melting temperature vs. size curve, as shown in Fig.~\ref{Fig.4.1}, asymtotically approaches a constant value of 5800 K $\pm$ 22 K, which is taken as an estimate of the melting temperature of free-standing infinite graphene. 

In the second experiment, in which prolonged-heating protocol is used, the melting temperature of infinitely long CNTs with various radius $R$ are obtained. The $R$-dependent melting temperature of CNT is as shown in Fig.~\ref{Fig.4.2}. The melting temperature of infinite size graphene, which is taken as the asymtotic value of $\lim_{R \to \infty} T_{\textit{SWCNT}}(R)$, is estimated as 5302 K $\pm$ 36 K. The melting temperature of infinite SWCNT converges to that of infinite size graphene when the radius $R > 1.15$ nm. The constant $C$ is estimated as 7.3 meV nm${}^{2}$. 

In the third experiment, in which prolonged-heating protocol is used, a melting temperature for infinite graphene of 5335 K $\pm$ 140 K is obtained, which is consistent with 5302 K $\pm$ 36 K from the second experiment but not the first. 

There is an apparent disparity between the results from the first experiment and that of the second and third experiments. The disparity may be due to the differences in the technical details in these MD simulations. The main differences between these experiments is that the finite graphenes in the first experiment have edges that are absent in the second and third experiments, which adopt periodic boundary condition. In addition, extrapolation made in the first experiment is based on finite graphenes of a largest size up to only 5000. The other difference is that in the second and third experiments, size convergence in the presence of periodic boundary condition has been performed, whereas conceptually we cannot perform such a size-convergence for finite graphenes in the first experiment. Based on the overall results of the MD experiments conducted, we have illustrated that estimating melting temperature of infinite graphene sheet by extrapolating the melting temperature of finite graphene sheets does not yield result consistent with that obtained with periodically constructed infinite graphene, probably because an infinite graphene sheet without any edge cannot be trivially mimicked by simply extrapolating the finite size to infinity, or at least, not until only 5000 atoms (which is the largest finite graphene size affordable by our present computational resource). We cautiously conclude that, based on the consistency of the data of the second and third experiments, that a free-standing infinite graphene sheet melts at the temperature of  
5302 K $\pm$ 36 K, using AIREBO forcefield.

\section{Acknowledgements}
T. L. Yoon wishes to acknowledge the support of (i) FRGS grant (Fasa 2/2013) by the Ministry of Higher Education of Malaysia (203/PfiZIK/6711348)l (ii) Universiti Sains Malaysia RU grant (No. 1001/PFIZIK/811240).

\bibliographystyle{model1-num-names}

\end{document}